# Negative Capacitance Enables FinFET Scaling Beyond 3nm Node

Ming-Yen Kao, Harshit Agarwal, Member, IEEE, Yu-Hung Liao, Suraj Cheema, Avirup Dasgupta, Member, IEEE, Pragya Kushwaha, Member, IEEE, Ava Tan, Sayeef Salahuddin, Fellow, IEEE, and Chenming Hu, Life Fellow, IEEE

*Abstract*—A comprehensive study of the scaling of negative capacitance FinFET (NC-FinFET) is conducted with TCAD. We show that the NC-FinFET can be scaled to "2.1nm node" and almost "1.5nm node" that comes two nodes after the industry "3nm node," which has 16nm $L_g$ and is the last FinFET node according to the International Roadmap for Devices and Systems (IRDS). In addition, for the intervening nodes, NC-FinFET can meet IRDS $I_{on}$ and $I_{off}$ target at target-beating $V_{DD}$. The benefits of negative capacitance (NC) include improved subthreshold slope (SS), drain-induced barrier lowering (DIBL), $V_t$ roll-off, transconductance over $I_d$ ($G_m/I_d$), output conductance over $I_d$ ($G_d/I_d$), and lower $V_{DD}$. Further scaling may be achieved by improving capacitance matching between ferroelectric (FE) and dielectric (DE).

*Index Terms*—Scaling, International Roadmap for Devices and Systems (IRDS), Landau equation, negative capacitance field-effect capacitance (NCFET), TCAD.[1]

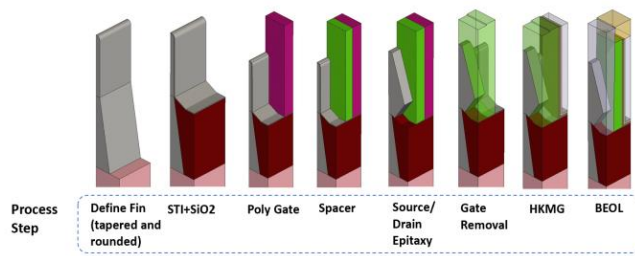

Fig. 1. Process simulation flow. Only the source side half of the FinFET is shown.

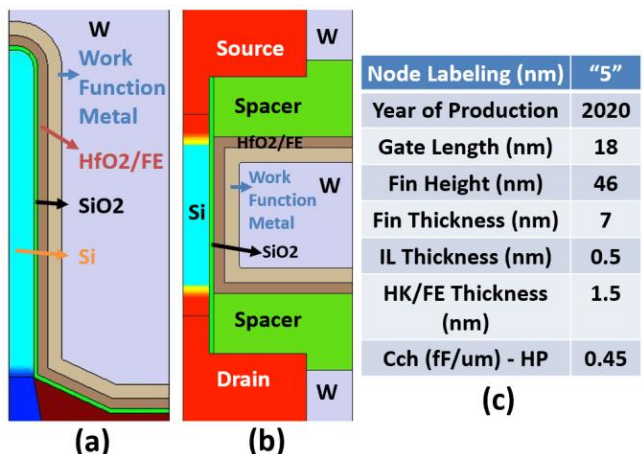

Fig. 2. (a) vertical and (b) horizontal cross sections of the half-FinFET (c) key geometry parameters of the simulated device. Note that the capacitance of the gate stack ($C_{ch}$) matches with IRDS high performance (HP) requirement, 0.45fF/$\mu$m.

## I. INTRODUCTION

The negative capacitance field-effect transistor (NCFET) is a promising technology for near future logic devices [1]. SS, DIBL, $V_t$ roll-off, $G_m/I_d$, and $G_d/I_d$ of the FinFET can be improved by doping Zr into the $HfO_2$ high-κ gate dielectric [1-2]. The show stopper to scaling of FinFET, according to IRDS, is the difficulty in reduction of fin-thickness ($T_{fin}$) and reduction of effective oxide thickness (EOT). Our study shows that NC enables FinFET scaling beyond "3nm node" without requiring further thinning of $T_{fin}$ and gate stack. The scalability of the metal-ferroelectric-metal-insulator-semiconductor (MFMIS) NCFET (with internal metal) has been discussed in [3]. Nevertheless, the electrical characteristics of the metal-ferroelectric-insulator-semiconductor (MFIS) NCFET (without internal metal) is different from the MFMIS NCFET [4], and an internal metal is not desirable in practical logic devices. In this paper, the scalability of MFIS NC-FinFET will be discussed. Moreover, the parameters of FE used in this paper are extracted from MFIS capacitor, not from the polarization-electric field (PE) loop of metal-ferroelectric-metal (MFM) structure. The extracted FE parameters are experimentally available and ready for MFIS NCFET.

## II. TCAD SIMULATION

Sentaurus Process Simulation [5] is used to build a realistic device for device simulations. The flow of process simulation is shown in Fig. 1. A gate-last process is adopted using a high-κ metal gate. Only the source side of the FinFET is shown for simplicity. Fig. 2 (a) and (b) show the cross-section of the fin. After the device structure is built, Sentaurus Device Simulation [6] is utilized to simulate the electrical characteristics. The electronic band structure with stress effects is included with carrier trajectory and scattering calculations and using the k.p. deformation potential model. Scattering mechanisms such as

[1] This work was supported by the Berkeley Device Modeling Center, University of California at Berkeley, CA 94720 USA. (Corresponding author: Ming-Yen Kao)

M.-Y. Kao, Y.-H. Liao, S. Cheema, A. Dasgupta, P. Kushwaha, A. Tan, S. Salahuddin, and C. Hu are with the Department of Electrical Engineering and Computer Sciences, University of California at Berkeley, Berkeley, CA 94720 USA (e-mail: mingyenkao@berkeley.edu).

H. Agarwal is with Indian Institute of Technology Jodhpur, Karwar, Rajasthan 342037, India.

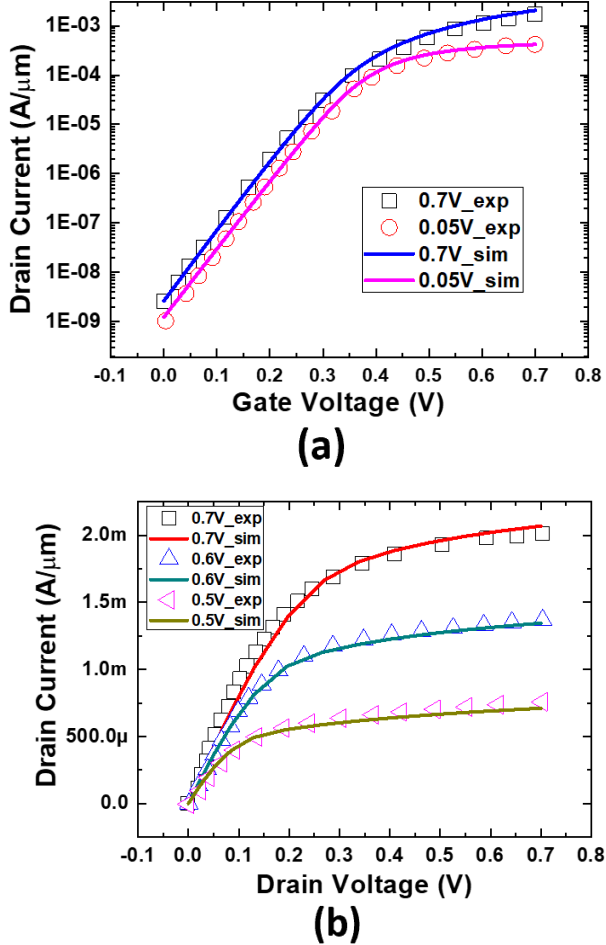

Fig. 3. TCAD FinFET calibration to 18nm $L_g$ Intel experimental data. *Per-foot-print* normalization of $I_d$ is done in this figure to match the Intel presentation. (a) $I_d$-$V_g$ fitting plot at high and low drain voltage (b) $I_d$-$V_d$ fitting plot at three different gate voltages, 0.7V, 0.6V, and 0.5V.

phonon scattering, impurity scattering, surface roughness scattering, remote Coulomb scattering, and impact ionization are also included. SRH and Auger recombination are additionally considered. Finally, drift-diffusion with the quantum confinement effect is solved self-consistently with the Sentaurus Device simulator.

An n-type FinFET with fin-height ($H_{fin}$) of 46nm, $T_{fin}$ of 7nm, and gate length ($L_g$) of 18nm is calibrated to the "Intel 10nm node" (equivalent to "5nm node" using the IRDS node definition) experimental data [7-8] as shown in Fig. 3. Note that the definition of *per-foot-print* (drain current ($I_d$) normalized to the fin pitch) is used only in Fig. 3 to match the Intel presentation [7], whereas all other figures and tables on $I_d$ in this paper present $I_d$ *per-channel-width* ($I_d$ normalized to the sum of 2 times $H_{fin}$ plus $T_{fin}$) because this work is not concerned with fin pitch. After this calibration, all TCAD parameters, including contact resistivity, are fixed, except the change of fin width from 7nm at "5nm node" to 6nm at "3nm node" according to the IRDS definition [8]. The gate stack is composed of a 0.5nm $SiO_2$ interfacial layer (IL) and either 1.5nm $HfO_2$ (FinFET) or 1.5nm HZO (NC-FinFET), the latter meaning that $HfO_2$ is doped with Zr and becomes FE in the NC-FinFET simulation. In NC-FinFET simulation, FE parameters, including α and β from Landau's Equation and background dielectric of FE ($\epsilon_{FE}$), are extracted from experimental C-V of NC MFIS structure [9] (see Fig. 4), and strength of the polarization gradient (domain coupling) is set to be $5 \times 10^{-5}$ $cm^3$/F (on the same order as [10]). In Fig. 4, gate insulator for both HK and NC constitutes of a chemical oxide (8Å) and 2.8nm layer of $HfO_2$ or HZO. The extracted dielectric constant of $HfO_2$ (HK) is 33 which corresponds to 1.1 nm EOT gate stack. On the other hand, very high dielectric constant (>100) exceeding theoretical predictions for Hf and Zr-based dielectrics [11-15] is required to fit the HZO C-V, and the anomalous I-V behavior in [9] must be explained by non-linear response of the gate insulator [16]. Therefore, a model with partially active FE layer in HZO is presented in [16] to explain both the C-V and I-V results in [9]. This work adopt the same methodology to use the Landau-Khalatnikov (L-K) model to extract HZO parameters and fit to Fig. 4, with $\alpha = -6.5 \times 10^{10} \frac{cm}{F}$ and $\beta = 8.1 \times 10^{19} \frac{cm^5}{FC^2}$.

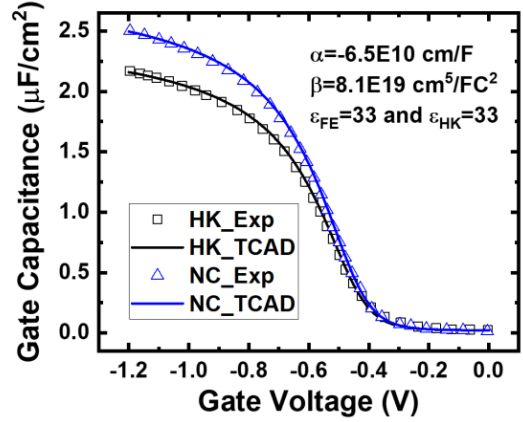

Fig. 4. C-V fitting of the experiment data of NC-MOSCAP. The extracted α and β are equivalent to $P_r$ = 20 μ $C/cm^2$ and $E_c$ = 1MV/cm.

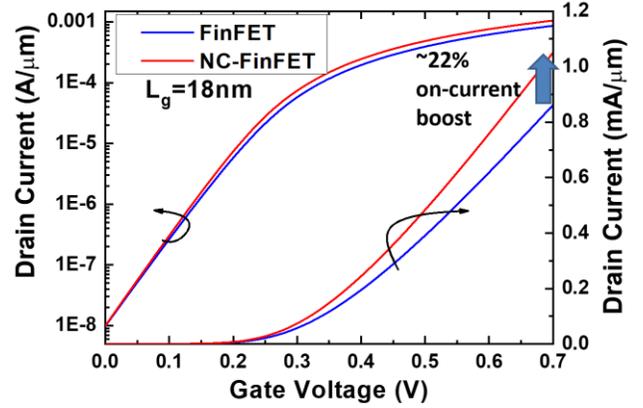

Fig. 5. $I_d$-$V_g$ plot of the FinFET and the NC-FinFET at $L_g$=18nm.

III. RESULTS AND DISCUSSION

Fig. 5 demonstrates the improvement from FinFET to NC-FinFET at $L_g$=18nm, and the boost of $I_{on}$ is about 22% when the $I_{off}$ is aligned at 10nA/μm. Table 1 shows the IRDS scaling targets from "5nm" to "1.5 nm". The second row shows the year of production [8]. The third row shows the physical gate length. Physical gate length is predicted to reach the scaling limit of 12nm according to IRDS. $T_{fin}$ (the fourth row of Table 1) is set to be 6nm from "3nm" to "1.5nm" according to IRDS definition. The 5th row in Table 1 shows IRDS target $V_{DD}$, and the 6th row shows the IRDS $I_{on}$ targets at $I_{off}$ = 10nA/$\mu$m. The 7th row shows that simulated FinFETs cannot meet the targets after "3nm node," hence the red color. The 8th row shows that NC-FinFETs with extracted FE parameters can meet IRDS requirement one more node beyond "3nm node" and almost meet "1.5nm node" with $I_{on}$ only 3% less than the target $I_{on}$.

| 1 | "2018 IRDS" Node | "5" | "3" | "2.1" | "1.5" |
|---|---|---|---|---|---|
| 2 | Year of Production | 2020 | 2022 | 2025 | 2028 |
| 3 | Physical Gate Length (nm) | 18 | 16 | 14 | 12 |
| 4 | Fin Width (nm) | 7 | 6 | | |
| 5 | IRDS Target VDD (V) | 0.70 | | 0.65 | |
| 6 | Ion Target (mA/μm) | 0.85 | 0.91 | 0.82 | 0.92 |
| 7 | Ion of FinFET (mA/μm) | 0.86 | 0.93 | 0.75 | 0.64 |
| 8 | Ion of NC-FinFET (mA/μm) | 1.05 | 1.12 | 0.93 | 0.89 |

Table 1. Simulation plan follows the IRDS 2018 roadmap. Red highlighting of the TCAD results indicate failure to meet the on-current targets at all future nodes.

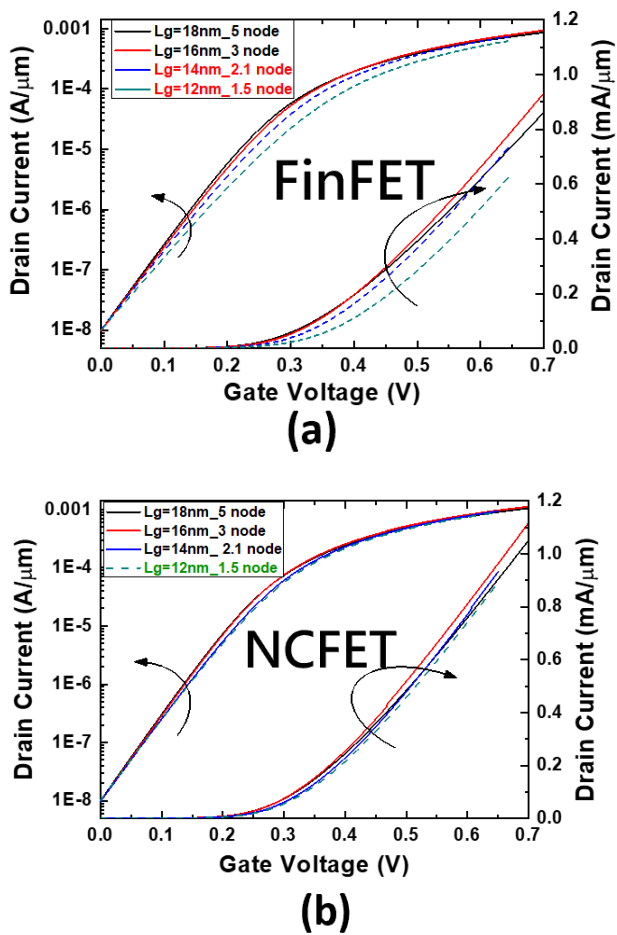

Fig. 6. $I_d$-$V_g$ of (a) FinFETs and (b) NC-FinFETs with work functions shifted to align the off-current with the IRDS high-performance requirement.

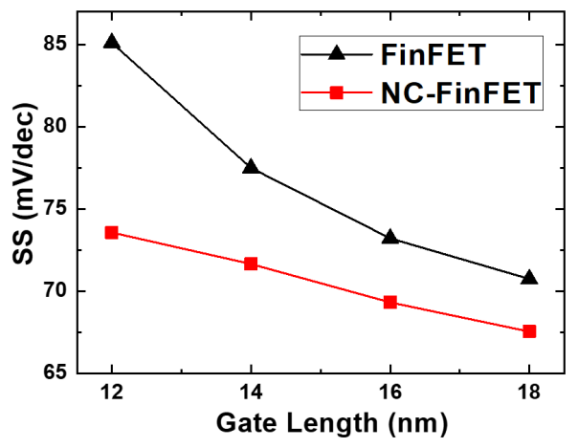

Fig. 7. SS of NC-FinFET is smaller than FinFET at 18nm $L_g$ and rises at lower rate with decreasing $L_g$.

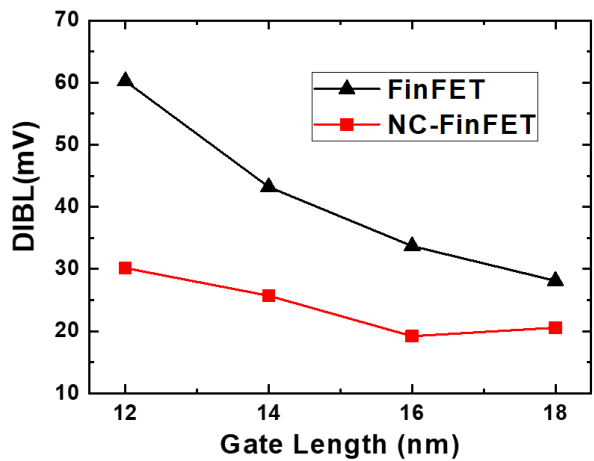

Fig. 8. DIBL versus gate lengths. DIBL is smaller in NC-FinFET and rises at lower rate as $L_g$ shrinks.

Fig. 6 shows $I_d$-$V_g$ simulation results at $V_d$ = IRDS $V_{DD}$ from "5nm node" to "1.5nm node" with work function shifted to align the $I_{off}$ at 10nA/$\mu$m. One can see that FinFETs beyond "3nm node" cannot meet the IRDS targets in Fig. 6 (a). The nodes which fail to reach the IRDS targets are labeled in red color and plotted in dash line. NC-FinFET, on the other hand, can meet the IRDS targets at "2.1nm node" and almost meet the IRDS target at "1.5nm node" respectively- two more nodes than FinFET in our simulations. The NC-FinFET simulation results are summarized in the 8rd row of Table 1. For several nodes, $I_{on}$ is significantly larger than the targets.

SS versus $L_g$ is shown in Fig. 7. SS degrades for both the NC-FinFET and FinFET when $L_g$ decreases, but the SS degrades at a lower rate for the NC-FinFET because the inner-fringing field, which becomes stronger at shorter $L_g$, helps capacitance matching and enhance $V_g$-amplification. Note that even if the SS of the NCFET is not below 60mV/dec in the "weak NC" FinFET studied here, the NC effect improves the on/off ratio improvement is large enough to enable 2 more nodes of scaling than simple FinFETs. Fig. 8 shows the DIBL versus different gate length of FinFETs and NC-FinFETs. NC helps relieve the degradation of DIBL, as it can be seen that the slower rate of increase in DIBL when gate length scales. Note that negative DIBL does not appear because the NC effect is "weak" (only 36% of the HZO is active FE layer) in this extracted parameter set. The $SS_{NC}/SS_{FinFET}$ and $DIBL_{NC}/DIBL_{FinFET}$ trend are consistent with Y. Liao et. al.'s results [17]. Fig. 9 compares the $I_d$-$V_d$ characteristics of the FinFETs and the NC-FinFETs from $L_g$=18nm to $L_g$=12nm. In Fig. 9, NC-FinFETs are plotted in solid lines, and FinFETs are plotted in dash lines. Current of FinFETs decreases a lot from $L_g$=18nm to $L_g$=12nm because of the short channel effect; whereas, current of NC-FinFETs barely decreases from $L_g$=18nm to $L_g$=12nm.

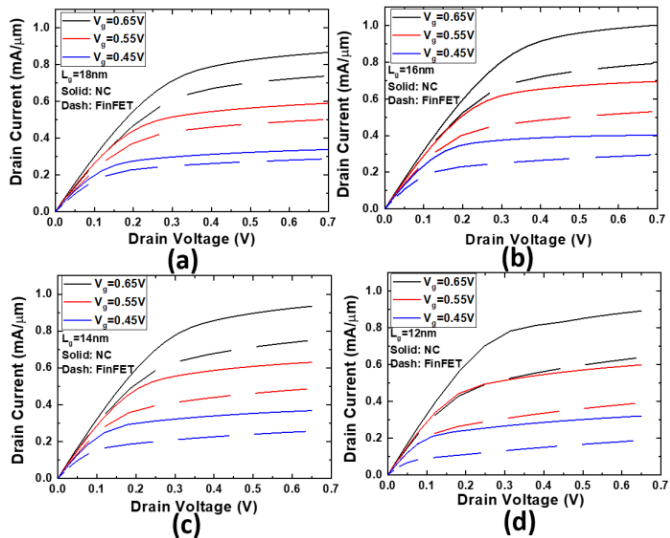

Fig. 9. $I_d$ is larger and $g_d$ is smaller in NC-FinFET than in FinFET at (a). $L_g$=18nm, at (b) $L_g$=16nm, at (c) $L_g$=14nm, and at (d) $L_g$=12nm.

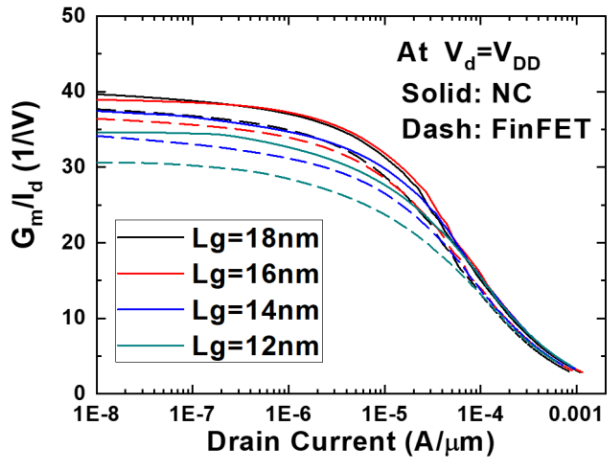

Fig. 10. $G_m/I_d$ versus drain current. NC-FinFETs have better $G_m/I_d$ performance than FinFETs overall.

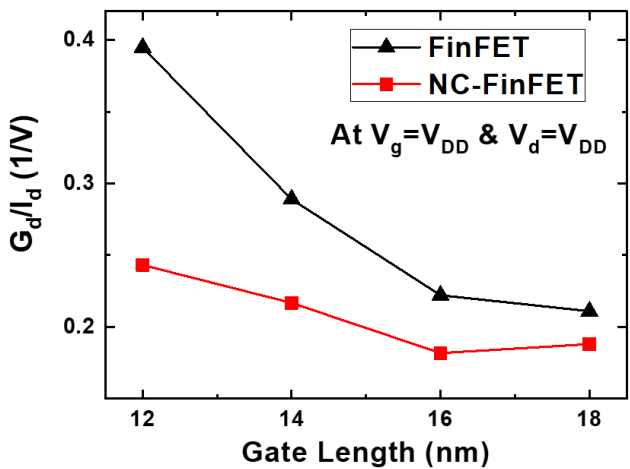

Fig. 11. $G_d/I_d$ versus drain current. NC-FinFET has higher $G_m/I_d$ and lower $G_d/I_d$ leading to better analog performance.

| "2018 IRDS" Node | "5" | "3" | "2.1" | "1.5" |
|---|---|---|---|---|
| Year of Production | 2020 | 2022 | 2025 | 2028 |
| IRDS Target VDD (V) | 0.70 | | 0.65 | |
| FinFET $V_{DD}$ needed to meet IRDS Target Ion | 0.70 | 0.69 | X | X |
| NC–FinFET $V_{DD}$ needed to meet IRDS Target Ion | 0.63 | 0.63 | 0.61 | X |

Table 2. $V_{DD}$ needed to reach IRDS target on-current ($I_{off}$ is fixed at 10nA/$\mu$m) for different IRDS nodes.

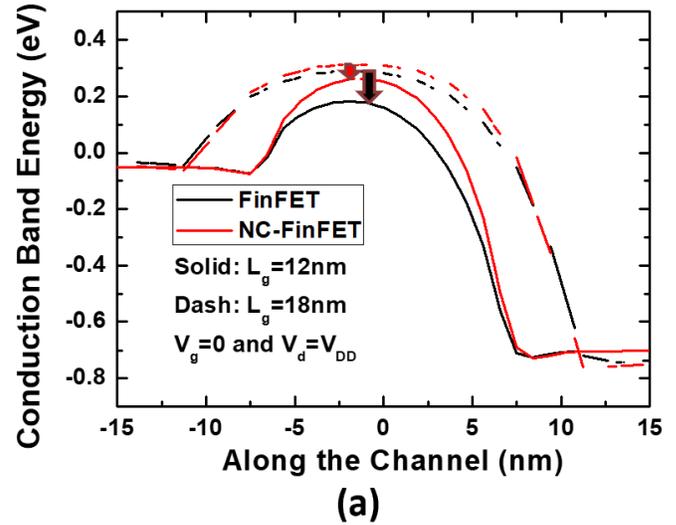

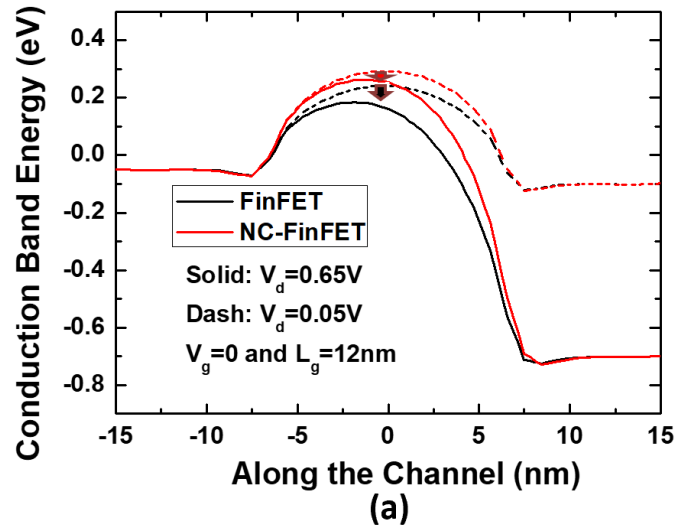

Fig. 12. (a) The potential barrier is higher in NC-FinFET (red) than in FinFET (black), and the difference is larger at Lg=12nm (solid) than at Lg=18nm (dash). (b) DIBL of NC-FinFET is smaller than FinFET. Note that the work function is not shifted in this figure.

The scaled NC-FinFETs are also good for analog applications. The $g_m/I_d$ versus $I_d$ of FinFET and NC-FinFET at $L_g$=18nm, 16nm, 14nm, and 12nm are presented in Fig. 10. The $g_m/I_d$ of the NC-FinFET is better than the conventional FinFET overall, and the $g_m/I_d$ of NC-FinFET at $L_g$=18nm and 16nm nearly hit the theoretical limits of 40 1/V at room temperature [18]. The $g_d$ to drain current ratio is shown in Fig. 11. While $g_d/I_d$ increases with shorter channel length for the FinFET due to the short channel effects, the trend of $g_d/I_d$ is the opposite for the NC-FinFET from $L_g$=18nm to $L_g$=16nm since the FE polarization induced by the inner-fringing field at short $L_g$ and high $V_d$ negates and overwhelms the short channel effect. For NC-FinFET at $L_g$=14nm and $L_g$=12nm, $g_d/I_d$ increase at much slower rate compared with FinFET. This benefits the intrinsic voltage gain, speed of both the static and pass-transistor logic, and noise margin of logic gates [19].

Table 2 shows another way to utilize NC-FinFETs' potential "excess horse power" in the intervening nodes. The $V_{DD}$ of the NC-FinFET is reduced for each $L_g$ by trial and error until $I_{on}$ and $I_{off}$ at $V_d$=reduced $V_{DD}$ match the IRDS $I_{on}$ and $I_{off}$ targets, respectively. The reduced (needed) $V_{DD}$ for the scaled NC-FinFET is shown in the last row of Table 2. Some IRDS $I_{off}$ and $I_{on}$ targets may be reachable below the target $V_{DD}$ by 70mV at significant power reduction.

Fig. 12 (a) shows the conduction band energy along the channel at $V_g$=0 and $V_d$=$V_{DD}$. The black arrow in Fig. 12 (a) demonstrates the reduction of the top-of-barrier (TOB) of FinFETs due to gate length scaling. In comparison, the red arrow in Fig. 12 (a) which indicates the reduction of the TOB in NC-FinFETs due to gate length scaling is much shorter than the black arrow. Fig. 12 (a) shows that NC-FinFETs have better immunity toward gate length scaling. Fig. 12(b) illustrates the reduced drain-induced barrier lowering effects at $L_g$=12nm. When the inner-fringing field, more significant at high $V_d$, goes through the channel to the FE film, it induces polarization in the FE such that the NC-FinFETs' channel potential barrier (the red line in Fig. 12 (b)) becomes higher compared with FinFETs' channel potential barrier (the black line in Fig. 12 (b)). That is why the reduction of the TOB due to the increase of drain bias in NC-FinFETs (the red arrow in Fig. 12 (b)) is smaller than the reduction of the TOB of FinFETs (the black arrow in Fig. 12 (b)).

## IV. CONCLUSION

NC may enable FinFET scaling 2 nodes beyond the "3nm node" without requiring thinner $W_{fin}$ or high-k film. We note that this is a TCAD simulation study that assumes the uniform FE film which can be scaled from large-area NC-MOSCAP to small NCFET device without changing the properties of the FE film and can be put into production. On the other hand, future HZO optimization with larger portion of active FE layer, multi-layer FEs [20], or varying FE along the channel [21], may lead to even much better NC performance in the future. NC may delay the need for nano-sheet FET in the near term and extend the nanosheet scalability in the long term.